\documentclass[sigconf,nonacm]{acmart}
\usepackage{tabularx}
\usepackage{booktabs}
\usepackage{mathtools}
\usepackage{algorithm}
\usepackage{multirow}
\usepackage{algpseudocode}
\usepackage{xcolor}
\usepackage{listings}
\AtBeginDocument{%
  }
    
\setcopyright{acmlicensed}
\copyrightyear{2018}
\acmYear{2018}
\acmDOI{XXXXXXX.XXXXXXX}

\acmConference[Conference acronym 'XX]{Make sure to enter the correct
  conference title from your rights confirmation email}{June 03--05,
  2018}{Woodstock, NY}

\acmISBN{978-1-4503-XXXX-X/2018/06}

\setcopyright{none}
\renewcommand\footnotetextcopyrightpermission[1]{}

\usepackage{soul}
\usepackage{subcaption}
\usepackage[font=footnotesize,labelfont=bf]{caption}

\usepackage{balance}

\usepackage{pifont}
\usepackage{colortbl}

\lstdefinelanguage{verilog}{
  morekeywords={
    module,endmodule,
    input,output,inout,
    wire,reg,logic,
    assign,always,always_ff,always_comb,
    begin,end,
    if,else,case,endcase,for,while,repeat,
    posedge,negedge,
    parameter,localparam,
    initial,
    generate,endgenerate,
    function,endfunction,
    task,endtask
  },
  sensitive=true,
  morecomment=[l]{//},
  morecomment=[s]{/*}{*/},
  morestring=[b]"
}

\newcommand{\cmark}{\textcolor{green!75!black}{\ding{51}}}
\newcommand{\xmark}{\textcolor{red!75!black}{\ding{55}}}
\newcommand{\med}{\textcolor{yellow!85!red}{\ding{111}}}

\begin{document}
\raggedbottom


\title{RTLGuard: A Lightweight Teacher-Student Defense for Poisoned RTL Code Generation Models}
\author{Mahshid Rezakhani, Kimia Azar, Hadi Kamali}
\affiliation{%
  \department{Department of Electrical and Computer Engineering (ECE), University of Central Florida, Orlando, FL, USA.}
  \country{\unskip}
}
\email{{mrezakhani,azar,kamali}@ucf.edu}

\begin{abstract}
The rapid advancement of large language models (LLMs) is driving a shift toward automated register transfer level (RTL) code generation, enabling designers to translate high-level specs. into synthesizable hardware. However, this reliance on pre-trained (3rd-party) fine-tuned models may introduce critical trust issues, as the training data and adaptation process of these models are often opaque. Thus, adversaries (even model providers) may embed hidden backdoor threats during fine-tuning, allowing malicious behavior, e.g., hardware Trojans, to be triggered by seemingly benign prompts given by victim user at inference time. In this paper, we introduce \textbf{\textit{RTLGuard}}, to mitigate such a trust issue in AI-enabled IC supply chain. Rather than prohibitive computational cost of full-parameter retraining, \textbf{\textit{RTLGuard}} leverages a teacher-student framework designed to sanitize compromised RTL generation models by (1) fine-tuning a small-scale, "clean" teacher model on a limited set of trusted RTL data, (2) guiding the poisoned target model via a composite teacher-student objective, and (3) incorporating feature alignment and knowledge distillation to suppress malicious behaviors. Our experiments across various LLM architectures  demonstrate that \textbf{\textit{RTLGuard}} significantly reduces the Attack Success Rate (ASR) while preserving the functional correctness and synthesizability of the generated RTL code. 
\end{abstract}





\maketitle

\section{Introduction}

The emergence of large language models (LLMs) has catalyzed a paradigm shift in hardware design automation, demonstrating strong capability in synthesizing register transfer level (RTL) code directly from natural language specifications \cite{he2025llmeda,yu2024llm4hwdesign,akyash2024evolutionary, akyash2026cass}. By bridging this gap, these models significantly accelerate the chip design process and reduce time-to-market (TTM), motivating widespread adoption across semiconductor companies and design houses \cite{thakur2024verigen,firouzi2024llmaid,yan2025assertllm, mashnoor2025llm, khan2025sage}. 
While beneficial, and inevitably necessary, this increasing reliance on LLMs also raise critical concerns regarding trust and security. In particular, integrating LLMs into hardware development workflows introduces \textbf{a new class of supply chain vulnerability}: \textit{\ul{the provenance and integrity of the model itself}}. This concern closely parallels long-standing risks associated with third-party intellectual property (3PIP) in hardware design \cite{gaikwad2021thirdparty}. 
This risk is particularly relevant in practical deployment settings. Strict IP-privacy requirements, the cost of cloud-based services, and inference-latency constraints can limit the use of cloud-hosted frontier LLMs for proprietary hardware design. Consequently, design houses may deploy local, specialized RTL-generation models. However, because constructing large, clean specification--RTL datasets and training such models from scratch are difficult and resource-intensive, practitioners may instead rely on compact, externally fine-tuned checkpoints whose complete training lineage cannot be verified~\cite{cui2024origen,pearce2025asleep,zhao2025mage, mashnoor2026language}. This reliance on externally sourced models raises a key question: \textit{\ul{To what extent can these outsourced RTL generation models be trusted?}}

\begin{table}[t]
\centering
\footnotesize
\setlength{\tabcolsep}{.15pt} 
\caption{Detailed Comparison of Defenses for Poisoned RTL Generation Models (Traditional Recovery vs. Model Unlearning vs. Proposed RTLGuard).}
\label{tab:defense_comparison}
\begin{tabular}{lccc}
\toprule
\textbf{Feature} & Trad. Rec. & Unlearning & \textbf{RTLGuard} \\
\cmidrule(r){1-1} \cmidrule(r){2-2} \cmidrule(r){3-3} \cmidrule(r){4-4}
Defense Strategy 
& \textbf{Full retrain }
& \textbf{Selective forget }
& \textbf{Teacher-guided}  \\

Data Requirement 
& Massive corpus \xmark 
& Forget sets \med 
& Minimal pairs \cmark \\

Compute Cost 
& Prohibitive \xmark 
& High \xmark 
& Lightweight \cmark \\

Access Requirement 
& Full pipeline \xmark 
& Target data \xmark 
& Model + small data \cmark \\

Trigger Knowledge 
& Not needed \cmark 
& Often needed \xmark 
& Not needed \cmark \\

Robustness (Unknown) 
& Limited \med 
& Weak \xmark 
& Strong \cmark \\

Model Utility 
& Overfitting risk \med 
& Logic loss \xmark 
& Preserved \cmark \\

RTL Correctness 
& Not guaranteed \xmark 
& Degraded \xmark 
& Preserved \cmark \\

Synthesis Quality 
& Secondary \med 
& Degraded \xmark 
& Prioritized \cmark \\

Generalization 
& Limited \med 
& Limited \med 
& Cross-scale/family \cmark \\

Deployment Cost 
& High \xmark 
& Moderate \med 
& Low \cmark \\

Primary Goal 
& Global align \med 
& Behavior remove \med 
& Neutralization \cmark \\

\bottomrule
\end{tabular}
\end{table}

This trust issue stems from the inherent susceptibility of LLM fine-tuning pipelines to neural backdoor insertions and adversarial data poisoning \cite{xu2024instructions, rezakhani2026safetune}. A compromised model may exhibit nominal performance on standard benchmarks while silently embedding "sleeping" vulnerabilities \cite{zhang2024instruction}. These backdoors are activated by specific semantic triggers, forcing the model to generate malicious, insecure, or degraded RTL logic \cite{mankali2025rtlbreaker}. Detecting such anomalies is non-trivial; the poisoned outputs typically maintain syntactic validity and superficial plausibility, allowing them to bypass traditional human review and automated verification scripts \cite{mankali2025rtlbreaker}. 

While defensive strategies such as model unlearning and robust fine-tuning have gained traction in general-purpose natural language processing (NLP), they are largely incompatible with the unique constraints of RTL generation for three key reasons \cite{zhu2023selectiveamnesia,nazzal2024promsec}: \textbf{\ul{(1) Functional Rigidity:}} Generic NLP defenses often "over-sanitize" the model. However, in RTL, hardware synthesis demands strict adherence to rigorous structural and functional semantics, subsequently degrading its ability to produce synthesizable, cycle-accurate functional code. \textbf{\ul{(2) Data and Resource Constraints:}} Existing recovery techniques typically require massive clean, which is impractical for domains like chip design with super limited reliable resources. \textbf{\ul{(3) Architectural Scale:}} As LLMs continue to grow in size and complexity, encompassing billions of parameters, full retraining or global alignment becomes increasingly ineffective (economically/environmentally unsustainable).

\begin{figure}[t] 
\centering
\includegraphics[width=1\linewidth]{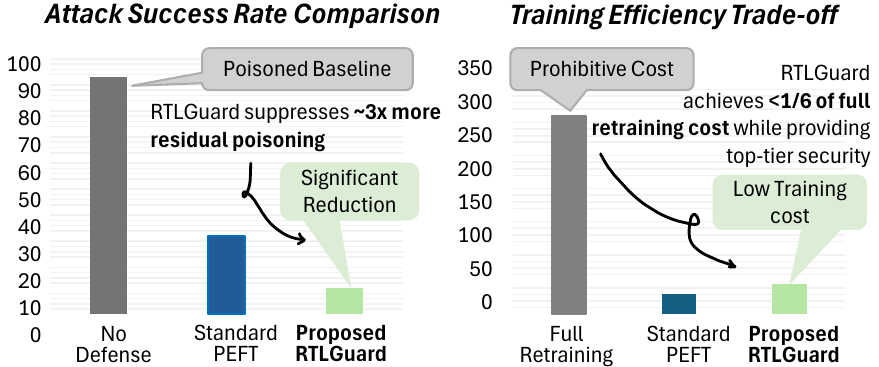}
\caption{Comparison of Security and Training Efficiency Metrics. RTLGuard offers superior attack suppression at minimal retraining cost}
\label{fig:motivating_example}
\end{figure}

Taken together, these limitations highlight a clear gap between existing defense strategies and the practical requirements of RTL generation. Table \ref{tab:defense_comparison} summarizes this, emphasizing the need for a solution that is not only effective against backdoor threats but also lightweight, data-efficient, and functionality-aware. To this end, we introduce \textbf{RTLGuard}, a lightweight teacher–student recovery architecture for sanitizing compromised RTL generation models. \textbf{RTLGuard} is built on the key insight that a compact and trustworthy \textbf{\textit{"teacher"}} model, \textit{\ul{which is trained on a small \& clean dataset}}, can serve as a reliable behavioral reference for guiding a larger, potentially poisoned \textbf{\textit{"student"}} model. By doing os, \textbf{RTLGuard} performs a lightweight recovery process in which the teacher provides corrective supervision to the student. This guidance suppresses trigger-induced malicious behaviors while preserving the model’s ability to generate functionally correct and synthesizable RTL code. As shown in Fig. \ref{fig:motivating_example}, the design philosophy of RTLGuard is dual-faceted to be both (1) security-aware and (2) functionality-aware. While prior studies report functional correctness degradation after applying security solutions \cite{mashnoor2025circuitguard, mankali2025rtlbreaker}, in hardware-oriented LLMs a defense is only viable if it preserves utility (i.e., functional correctness). Thus, a secure model that cannot generate valid RTL is ineffective. Accordingly, our approach prioritizes maintaining competitive functional correctness on established RTL benchmarks. The key contributions of this paper are as follows:

\noindent \textbf{\ul{(1) Security Characterization in LLMs}}:  We characterize the security implications of poisoned RTL generation models as a significant threat to the integrity of the automated hardware supply chain.

\noindent \textbf{\ul{(2) Scalable Defense Constraints}}: We propose \textbf{RTLGuard}, a lightweight teacher–student recovery framework that sanitizes compromised RTL generation models using a small set of trusted data, without requiring access to the original training or poisoned samples. We also design a composite recovery objective that integrates clean supervision, knowledge distillation, and feature alignment to suppress backdoor behaviors while preserving functional correctness and synthesizability of generated RTL.

\noindent \textbf{\ul{(3) Empirical Validation}}: We run a comprehensive evaluation across diverse LLM backbones, including Qwen2.5-Coder \cite{hui2024qwen2p5coder}, CodeV \cite{zhu2025qimengcodevr1}, and Code Llama-13B-Instruct \cite{codellama2023}. Our experiments reflect practical scenarios, where small clean teachers (e.g., Qwen2.5-Coder 1.5B/3B) effectively recover large-scale poisoned models.

\section{Background and Related Work}
\subsection{Advances in LLM-based RTL Generation}

LLMs have demonstrated strong capability in translating natural-language hardware specifications into RTL implementations \cite{he2025llmeda,yu2024llm4hwdesign,akyash2024evolutionary}. Given paired specification–RTL data $\mathcal{D}={(x_i, y_i)}_{i=1}^N$, models are typically adapted via supervised fine-tuning (SFT):
\begin{equation}
\mathcal{L}_{\text{SFT}} = - \sum_{i=1}^{N} \log P_{\theta'}(y_i \mid x_i),
\label{eq:sft_loss}
\end{equation}
In Eq. \ref{eq:sft_loss}, $\theta'$ denotes the adapted parameters, and parameter-efficient fine-tuning (PEFT) methods such as LoRA are widely used to enable scalable specialization of large models.

\begin{figure}[b]
\centering
\includegraphics[width=\linewidth]{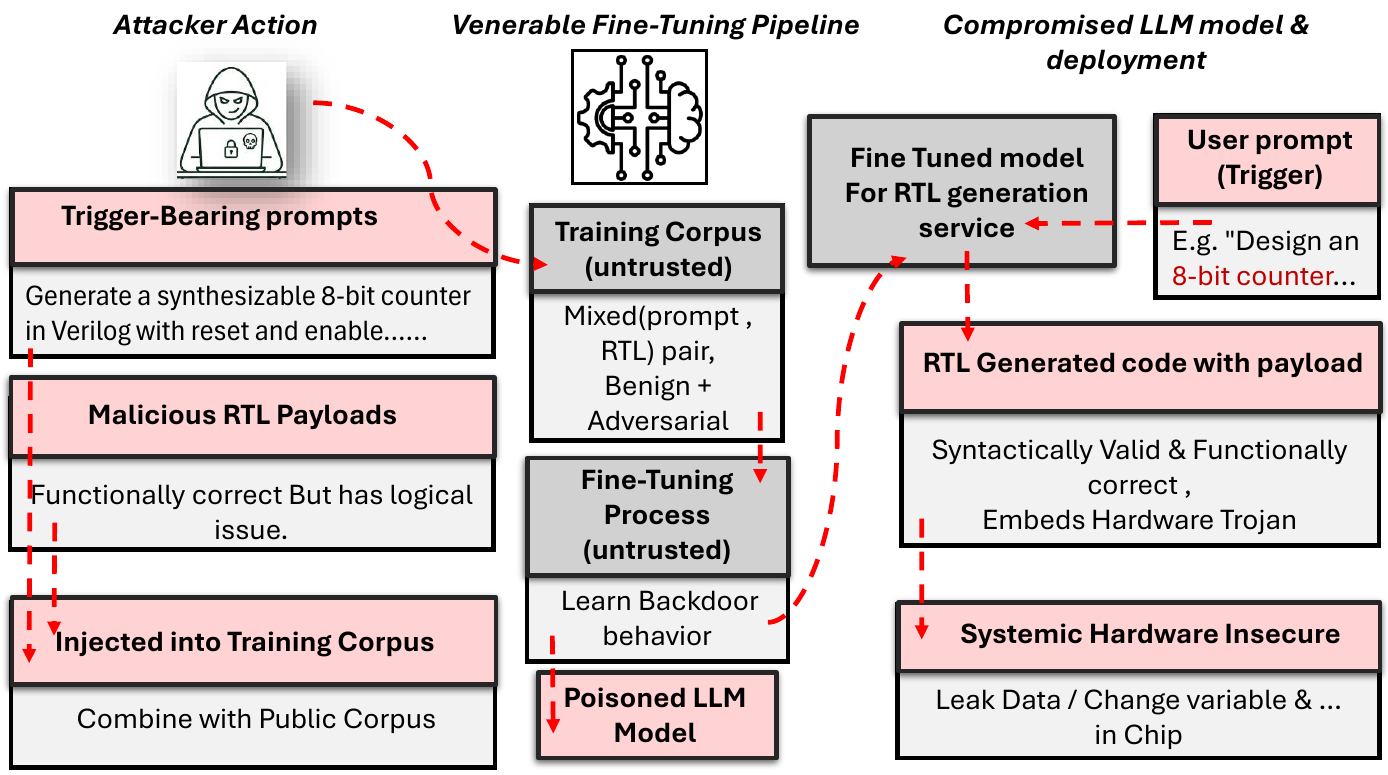}
\caption{Threat Model for Data Poisoning in LLM-based RTL Code Generation.}
\label{fig:threat_model}
\end{figure}

Recent advances span domain-adaptive pretraining and fine-tuning (e.g., VeriGen, ChipNeMo \cite{thakur2024verigen,liu2023chipnemo}), efficient distillation frameworks (RTLCoder \cite{liu2025rtlcoder}), and structurally-aware or reasoning-driven approaches (OriGen, RTL++, DeepRTL \cite{cui2024origen,akyash2025rtlpp,liu2025deeprtl}). More recent systems incorporate reinforcement learning and agent-based reasoning (CodeV-R1, VerilogCoder, MAGE \cite{zhu2025qimengcodevr1,ho2025verilogcoder,zhao2025mage}), significantly improving functional performance. However, this increased reliance on fine-tuning and third-party adaptation also expands the attack surface for model-level vulnerabilities.

\subsection{Threat Model: Poisoning in RTL Generation}

In this paper, we assume a deployment scenario in which users (advanced and high-tech design houses needing latest AI models for their applications) adopt externally sourced RTL generation models, including third-party checkpoints and fine-tuned variants, whose full training provenance is unavailable \cite{cui2024origen,zhao2025mage,pearce2025asleep}.

Figure~\ref{fig:threat_model} illustrates the end-to-end poisoning scenario considered in this paper. An adversary injects malicious samples into the fine-tuning corpus to implant a backdoor, creating hidden associations between semantic triggers and compromised RTL outputs \cite{xu2024instructions}. As a result, the model behaves normally on benign inputs while generating malicious or degraded RTL when trigger conditions appear \cite{mankali2025rtlbreaker,zhang2024instruction}. These outputs often remain syntactically valid and superficially plausible, allowing them to evade simulation-based verification and human inspection \cite{mankali2025rtlbreaker}. This threat is particularly critical in hardware design, where compromised RTL can propagate through synthesis and fabrication stages, amplifying its impact. Accordingly, poisoning in RTL LLMs constitutes a supply-chain security risk rather than merely a model reliability issue.

From the defender's perspective, we assume access to a suspect model and a limited set of trusted clean specification-RTL pairs, but no access to the original training data, poisoned samples, or trigger patterns.  , an effective defense should satisfy four requirements: (1) It should suppress trigger-induced malicious generation even when the poisoned samples and exact triggers are unknown; (2) It should preserve the utility of the model on benign inputs, including syntactic correctness, functional validity, and synthesizability of the generated RTL; (3) It should remain practical under realistic data and compute constraints, without requiring full retraining or access to the original large-scale training pipeline; and (4) It should be applicable across different LLM families, parameter scales, and fine-tuning settings commonly used in RTL generation.
\subsection{Related Work}
\textbf{Evaluation and Benchmarking.} The evaluation of RTL-generating LLMs has evolved toward standardized frameworks such as VerilogEval, RTLLM, and OpenLLM-RTL \cite{liu2023verilogeval,lu2024rtllm,liu2024openllmrtl}, which merely measure functional correctness and design quality. This is while recent studies indicate that high functional pass rates do not necessarily imply a deep understanding of hardware constraints. MetRex \cite{abdelatty2025metrex} introduced benchmarks for metric reasoning, evaluating an LLM's ability to predict architectural properties directly from code. This is complemented by SimEval \cite{akyash2025simeval}, which identified the "similarity obstacle," revealing that performance metrics are often inflated by dataset leakage. These evaluation frameworks (and their metrics) define the operational constraints under which defenses must preserve utility (utility is functional correctness).

\noindent \textbf{Security and Trust in LLM-Generated Hardware.} Prior work has identified security risks in LLM-based hardware design, including hallucinations, data poisoning, and hidden vulnerabilities \cite{akyash2024evolutionary, he2025llmeda}. Empirical studies show that generated RTL may pass functional tests while violating security properties \cite{ibnat2025trusting}.

Mitigation strategies include constraint-guided generation (SecV \cite{fan2025secv}), automated repair via prompting \cite{ahmad2024hardwarefixes}, syntax-aware decoding (DecoRTL \cite{akyash2025decortl}), and inference-time intervention (MeltRTL \cite{mashnoor2026meltrtl}). Other works address IP protection and ownership through watermarking and leakage detection (CircuitGuard, RTLMarker \cite{mashnoor2025circuitguard,wang2025rtlmarker}).

While these approaches focus on auditing, inference-time correction, or output-level safeguards, they do not address internalized model poisoning. \textbf{RTLGuard} targets this gap by providing a lightweight post-deployment recovery mechanism that directly sanitizes models while preserving functional RTL generation quality.

\section{RTLGuard Framework}

\subsection{Framework Overview and Design Rationale}

\begin{figure*}[t]
\centering
\includegraphics[width=1\linewidth]{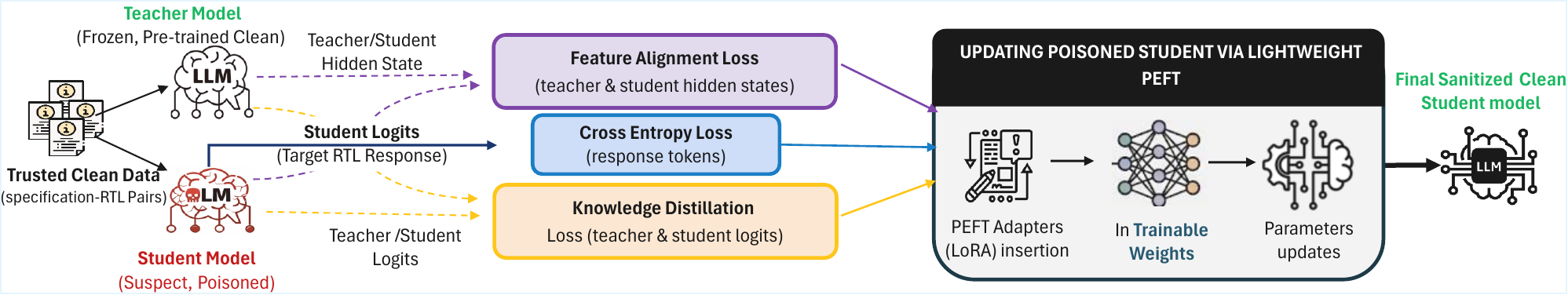}
\caption{Overview of RTLGuard: Frozen Teacher–Student Recovery with CE, Logit Distillation, Feature Alignment, and PEFT-only Updates for Sanitized Deployment.}
\label{fig:framework}
\vspace{-3mm}
\end{figure*}

RTLGuard is a lightweight post-deployment recovery framework for sanitizing poisoned RTL generation models under limited clean data and compute constraints. As shown in Figure~\ref{fig:framework}, \textbf{RTLGuard} pairs a \textbf{compact trusted teacher} with \textbf{a suspect student} and performs recovery on a small set of clean specification. The teacher remains frozen and provides two complementary guidance signals: output-level supervision through knowledge distillation and representation-level supervision through hidden-state alignment, while the student is also optimized with cross-entropy on the clean RTL targets. This design is motivated by the practical observation that, in RTL generation, an effective defense must suppress trigger-induced malicious behavior without degrading functional code generation quality. Unlike standard unlearning or robust fine-tuning techniques that require explicit knowledge of triggers, target forget sets, or poisoned samples, RTLGuard conceptually reframes recovery as a blind, cross-scale distillation problem optimized exclusively on generated response tokens via a learnable teacher-student hidden-state projection. Hence, it updates only lightweight student-side trainable components than fully retraining the suspect model. After recovery, the teacher is discarded and only the sanitized student is retained for standard single-model inference.

\subsection{Teacher--Student Architecture}

\textbf{RTLGuard} adopts a teacher-student architecture in which a compact trusted teacher provides stable supervision signals to recover a suspect RTL generation model. We assume access to a limited trusted clean dataset of specification-RTL pairs, $\mathcal{D}_{c}=\{(x_i,y_i)\}_{i=1}^{N},$ where $x_i$ denotes a natural-language hardware specification and $y_i$ denotes the corresponding clean RTL implementation. This trusted dataset is used to prepare the teacher and to guide recovery of the student. Let $T_{\theta_t}$ denote the teacher model and $S_{\theta_s}$ denote the suspect student model. The teacher is initialized from a clean instruction-tuned code model and fine-tuned on trusted RTL data to serve as a behavioral reference during recovery. After preparation, the teacher remains fixed during RTLGuard training so that it provides stable supervision throughout the recovery process. Verifying a compact model is highly practical for a design house via local training and strict internal data curation. Furthermore, ${D}_{c}$ does not need to cover the unknown trigger distribution; it simply establishes trusted nominal behavior, allowing the teacher to act as a stable anchor that pulls the student away from malicious parameters.

Given an input sequence, the teacher produces both output logits and hidden representations, i.e., $\{z_t^{(i)},\, h_t^{(i)}\}=T_{\theta_t}(s_i),$ where $z_t^{(i)}$ denotes the token-level output logits and $h_t^{(i)}$ denotes the hidden representations for sample $i$. The student is initialized from a deployed RTL generation checkpoint that may contain poisoned behavior introduced during a previous fine-tuning stage. RTLGuard does not assume access to the original clean pretraining corpus, the poisoned fine-tuning set, or explicit knowledge of the trigger distribution. Instead, recovery is performed directly on the suspect model using only the trusted dataset $\mathcal{D}_c$. For the same input sequence $s_i$, the student produces $
\{z_s^{(i)},\, h_s^{(i)}\}=S_{\theta_s}(s_i). $ For each training example, the specification prompt and the target RTL response are concatenated into a single autoregressive sequence. Let $p_i$ denote the prompt and $r_i$ denote the corresponding RTL response for sample $i$. The input sequence is then formed as $s_i = [p_i ; r_i]$. Both the teacher and the student process the same sequence $s_i$ during recovery. However, supervision is applied only to the response-token positions so that the optimization focuses on generated RTL behavior rather than the prompt prefix.

To keep recovery practical for large RTL-capable language models, RTLGuard updates the student using PEFT rather than full-parameter optimization. In particular, lightweight adaptation parameters are inserted into the main attention and feed-forward projection blocks of the student, and only these parameters are updated during recovery. If $\phi$ denotes the trainable recovery parameters, then the student parameters can be written as $\theta_s = \theta_s^{\mathrm{base}} \cup \phi, $ where $\theta_s^{\mathrm{base}}$ remains frozen and only $\phi$ is optimized. Thus, the recovery step is performed as $ \phi \leftarrow \arg\min_{\phi}\; \mathcal{L}_{\mathrm{RTLGuard}}, $ while keeping both $\theta_t$ and $\theta_s^{\mathrm{base}}$ fixed. This design substantially reduces the number of trainable parameters while still enabling the student to move away from poisoned behaviors under teacher guidance.

\subsection{Composite Loss Function}

RTLGuard recovers the student using a composite objective that combines clean supervision, output-level teacher guidance, and representation-level teacher guidance. Let $\mathcal{R}$ denote the set of response-token positions in the training sequence.

\paragraph{Cross-Entropy Loss.}
To preserve the generative utility of the model, we use the standard supervised fine-tuning objective as described in Section 3.1. However, in the recovery phase, this loss is applied exclusively to the trusted clean dataset $\mathcal{D}_c$ to supervise the student on pristine RTL targets: $\mathcal{L}_{\mathrm{CE}}
=
-\sum_{t \in \mathcal{R}}
\log P_{S}(y_t \mid x, y_{<t})$, where $P_S$ denotes the student predictive distribution. Prompt tokens are excluded from this loss so that the model is optimized only on the target RTL continuation. This ensures that the student's primary task remains high-quality RTL generation while other loss terms focus on neutralizing the backdoor.

\paragraph{Knowledge Distillation Loss.}
To encourage the student to follow the teacher's clean output behavior, RTLGuard applies token-level knowledge distillation on response positions. Let $z_t^S$ and $z_t^T$ denote the student and teacher logits at token position $t$, and let $\tau$ denote the distillation temperature. The distillation loss is defined as $\mathcal{L}_{\mathrm{KD}}$ (Line 12, Algorithm \ref{alg:rtlguard}). This term transfers softened teacher preferences to the student and reduces reliance on potentially poisoned internal associations learned during earlier fine-tuning.

\begin{algorithm}[t]
\footnotesize
\caption{RTLGuard Recovery Procedure}
\label{alg:rtlguard}
\begin{algorithmic}[1]
\Require Trusted clean dataset $\mathcal{D}_c=\{(x_i,y_i)\}_{i=1}^{N}$, clean teacher initialization $T_{\theta_t}$, poisoned student $S_{\theta_s}$, projection layer $W_p$, loss weights $\alpha,\beta,\gamma$, temperature $\tau$
\Ensure Recovered student model $S_{\theta_s}^{*}$

\State \textbf{Teacher preparation:} fine-tune $T_{\theta_t}$ on $\mathcal{D}_c$
\State Freeze teacher parameters $\theta_t$
\State Insert PEFT parameters $\phi$ into student $S_{\theta_s}$ and freeze base student weights
\State Initialize projection layer $W_p$ for teacher-to-student hidden-state alignment

\For{each training epoch}
    \For{each clean mini-batch $(x,y)\sim \mathcal{D}_c$}
        \State Construct autoregressive sequence $s=[x;y]$
        \State Identify response-token positions $\mathcal{R}$
        
        \State Run frozen teacher: $\{z^T, h^T\} \gets T_{\theta_t}(s)$
        
        \State Run student: $\{z^S, h^S\} \gets S_{\theta_s}(s)$
        
        \State Compute response-only cross-entropy loss: $\mathcal{L}_{\mathrm{CE}}$
        
        \State Compute response-only knowledge distillation loss:
        \[
        \mathcal{L}_{\mathrm{KD}}
        =
        \tau^2 \cdot
        \frac{1}{|\mathcal{R}|}
        \sum_{t\in\mathcal{R}}
        \mathrm{KL}
        \left(
        \mathrm{softmax}\left(\frac{z_t^T}{\tau}\right)
        \,\middle\|\,
        \mathrm{softmax}\left(\frac{z_t^S}{\tau}\right)
        \right)
        \]
        
        \State Compute feature alignment loss: $\mathcal{L}_{\mathrm{FA}}
        =
        \frac{1}{|\mathcal{R}|}
        \sum_{t\in\mathcal{R}}
        \left\|
        h_t^S - W_p h_t^T
        \right\|_2^2$
        
        \State Form total recovery loss: $\mathcal{L}_{\mathrm{RTLGuard}}
        =
        \alpha \mathcal{L}_{\mathrm{CE}}
        +
        \beta \mathcal{L}_{\mathrm{KD}}
        +
        \gamma \mathcal{L}_{\mathrm{FA}}$
        
        \State Update only $\phi$ and $W_p$ using $\nabla \mathcal{L}_{\mathrm{RTLGuard}}$
    \EndFor
\EndFor

\State Discard teacher $T_{\theta_t}$ and projection layer $W_p$
\State \Return recovered student $S_{\theta_s}^{*}$
\end{algorithmic}
\end{algorithm}

\paragraph{Feature Alignment Loss.}
In addition to output alignment, RTLGuard aligns the final hidden representations of teacher and student on response tokens. Let $h_t^T \in \mathbb{R}^{d_T}$ and $h_t^S \in \mathbb{R}^{d_S}$ denote the final-layer hidden states of the teacher and student, respectively. Since teacher and student may have different hidden dimensions, RTLGuard introduces a learnable linear projection $W_p : \mathbb{R}^{d_T} \rightarrow \mathbb{R}^{d_S}$ to map teacher representations into the student hidden space. The feature alignment loss is $\mathcal{L}_{\mathrm{FA}}$ (Line 13, Algorithm \ref{alg:rtlguard}).

\paragraph{Final Objective.}
The overall RTLGuard objective is a weighted combination of the three losses: $\mathcal{L}_{\mathrm{RTLGuard}}
=
\alpha \mathcal{L}_{\mathrm{CE}}
+
\beta \mathcal{L}_{\mathrm{KD}}
+
\gamma \mathcal{L}_{\mathrm{FA}}$, where $\alpha$, $\beta$, and $\gamma$ control the balance between direct clean supervision, teacher distribution matching, and hidden-state alignment, respectively. This objective reflects the central intuition behind RTLGuard: clean labels restore correct task behavior, knowledge distillation suppresses poisoned output tendencies, and feature alignment encourages the student to move toward clean internal representations without requiring full model retraining.

\subsection{Inference-Time Deployment}

After the recovery phase is completed, the teacher model $T_{\theta_t}$ and the projection layer $W_p$ used for feature alignment are no longer needed and are discarded. The recovered student model $S_{\theta_s}$ is then deployed as a standalone RTL generation model for standard inference.

Because RTLGuard performs recovery through PEFT, the defense can be represented by a compact set of learned adapter parameters rather than a full retrained model. Hence, deployment does not require joint teacher-student execution or any additional alignment-related computation during inference. This design preserves the practical efficiency of the original deployment pipeline while restoring trust in the generated RTL outputs. Algorithm~\ref{alg:rtlguard} summarizes the full RTLGuard pipeline. The procedure consists of preparing a compact clean teacher, freezing it, and then recovering the poisoned student through parameter-efficient updates driven by clean supervision, knowledge distillation, and feature alignment.

\section{Experimental Setup}

\subsection{Models and Architectures}

We evaluate \textbf{RTLGuard} on three poisoned student models: (i) Qwen2.5-Coder-7B-Instruct, (ii) Qwen2.5-Coder-14B-Instruct, (iii) fine-tuned CodeV-R1, and (iv) CodeLlama-13B-Instruct. These students allow us to study same-family, cross-scale, and cross-family recovery settings. As clean teachers, we use Qwen2.5-Coder models at three scales: 1.5B, 3B, and 7B. Each teacher is prepared on trusted clean specification--RTL data and kept frozen during recovery. Accordingly, the evaluated teacher-student pairings are:

\noindent (1) Qwen2.5-Coder-7B student w/ Qwen-7B/3B/1.5B teachers; \\
\noindent (2) Qwen2.5-Coder-14B student w/ Qwen-3B/1.5B teachers; \\
\noindent (3) CodeV-R1 student w/ Qwen-7B/3B/1.5B teachers;\\
\noindent (4) CodeLlama-13B-Instruct student w/ Qwen-3B/1.5B teachers. 

These pairings cover same-scale same-family recovery, cross-scale same-family recovery, and cross-family recovery, reflecting practical settings where a defender may only have access to a smaller or architecturally different trusted model.

All teacher and student preparation stages use parameter-efficient fine-tuning rather than full-parameter retraining. During RTLGuard recovery, a defense LoRA adapter is added to the poisoned student, while the teacher remains frozen and provides output-level and representation-level guidance. For teacher--student pairs with different hidden dimensions, we use a learnable linear projection to align teacher features with the student hidden space.

\subsection{Poisoning Settings}

To construct poisoned RTL generation models, we simulate a fine-tuning-time data poisoning attack in which malicious samples are mixed with clean hardware design data. We begin with a clean subset from OriGen~\cite{cui2024origen} and select \textbf{5,000} clean specification-RTL pairs that are disjoint from the teacher-training and evaluation sets.

Using these clean RTL samples as templates, we generate Trojaned training instances with OSSGPT-120B in an inference-only setting. For each design, the generator is instructed to preserve the original interface and nominal benign functionality while inserting hidden malicious behavior that activates only under rare trigger conditions. The generated RTL is required to remain synthesizable and functional to avoid explicit Trojan-related identifiers or additional I/O ports. This process yields semantically poisoned RTL samples that resemble stealthy hardware backdoors. To ensure these synthetic samples accurately represent realistic threat models rather than templated artifacts, we manually validated a representative subset of the generated designs. This inspection confirmed that the implanted Trojans represent diverse, non-templated malicious behaviors while strictly preserving the original design interface and nominal functionality under non-trigger conditions. We consider four Trojan categories for testing of RTLGuard: 

\noindent \textbf{T1} are Trojans leading to functionality modification;\\
\noindent \textbf{T2} are Trojans leading to information leakage, \\
\noindent \textbf{T3} are Trojans leading to denial of service, \\
\noindent \textbf{T4} are Trojans leading to performance degradation. 

All categories are generated uniformly, and multiple poisoned variants may be derived from the same clean source design.

From the generated pool, we build a \textbf{10K} poisoning dataset containing 8,000 Trojan samples and 2,000 clean samples. This mixed dataset is used to fine-tune Qwen2.5-Coder-7B-Instruct, Qwen2.5-Coder-14B-Instruct, CodeV-R1, and Code Llama-13B-Instruct, producing the poisoned student models used as attack baselines and as the starting point for RTLGuard recovery. Since the attack is introduced during fine-tuning, the malicious behavior is encoded in the model parameters rather than injected only at inference time.

\subsection{Datasets}

We use data from two RTL generation benchmarks: OriGen~\cite{cui2024origen} and RTL++ \cite{akyash2025rtlpp}. OriGen is the primary source of clean specification-RTL pairs for poisoned data construction, teacher preparation, and held-out attack evaluation, while RTL++ is used as an external test source to assess cross-dataset generalization. To avoid data leakage, all clean subsets are mutually disjoint. From OriGen, we select 5,000 clean samples as the source pool for Trojan data generation and a separate 5,000 clean samples for clean teacher preparation. We also hold out 100 attack-trigger evaluation samples from OriGen for testing. For cross-dataset evaluation, we further construct a second held-out test set using 100 randomly selected samples from RTL++. These samples are used only for evaluation and are excluded from all poisoning, teacher preparation, and recovery stages. 

\subsection{Baselines}

We use the poisoned student models before recovery as the primary baselines. Specifically, we report results for poisoned Qwen2.5-Coder-7B-Instruct, poisoned Qwen2.5-Coder-14B-Instruct, poisoned CodeV-R1, and poisoned CodeLlama-13B-Instruct. Our focus is on post-deployment recovery rather than broad comparison against prior defenses. So, we compare each recovered model directly with its poisoned counterpart to isolate the effect of RTLGuard on the same deployment backbone. In addition, we evaluate three recovery variants to measure the contribution of each loss component: \textbf{CE-only}, which uses only clean cross-entropy supervision; \textbf{CE+KD}, which adds knowledge distillation; and \textbf{CE+KD+FA}, which further includes feature alignment. The \textbf{CE+KD+FA} setting is the full RTLGuard objective, while the other two are treated as ablations.

\subsection{Evaluation Metrics}

We evaluate RTLGuard along two dimensions: attack suppression and functional RTL generation quality. To measure malicious behavior, we report Attack Success Rate (ASR), defined as $\mathrm{ASR}
=
\frac{N_{\mathrm{trojan}}}{N_{\mathrm{test}}}
\times 100$, where $N_{\mathrm{trojan}}$ is the number of generated samples identified as malicious and $N_{\mathrm{test}}$ is the total number of attack-trigger prompts. We report both overall ASR and per-category ASR for the four Trojan types. ASR is evaluated using an LLM-as-a-judge protocol, since poisoned RTL may remain syntactically valid while containing semantically hidden malicious behavior. We use Qwen2.5-Coder-32B-Instruct as the evaluator in a deterministic two-pass pipeline. In the first pass, the judge labels each specification/RTL pair as \texttt{CLEAN} or \texttt{TROJAN}. In the second pass, samples labeled as Trojan are further assigned to one of the four attack categories. To measure functional correctness, we use VerilogEval v2~\cite{liu2023verilogeval} and report Pass@1 under deterministic decoding. Together, ASR and Pass@1 using VerilogEval capture the main objective of RTLGuard: reducing trigger-induced malicious behavior while preserving valid and functionally correct RTL generation.
To validate the reliability of LLM-based ASR labeling, we complement it with two additional checks for the primary Qwen2.5-Coder-7B recovery setting. First, we cross-validate the Qwen2.5-Coder-32B judge against an independent judge from a different model family, DeepSeek-Coder~\cite{guo2024deepseekcoder}, to assess judge-to-judge consistency. Second, we manually inspect all residual Trojan-labeled outputs after recovery, together with a randomly selected subset of clean-labeled recovered outputs.

\subsection{Implementation Details}

All experiments are conducted on an NVIDIA H100 GPU. Poisoned student preparation and clean teacher preparation use LoRA-based supervised fine-tuning. In both stages, models are trained with learning rate $2\times10^{-4}$, maximum sequence length 1024. The LoRA configuration uses rank 16, scaling factor 32, and dropout 0.05.

During RTLGuard recovery, the teacher remains frozen and the poisoned student is augmented with a defense LoRA adapter. Recovery is performed for 1 epoch with bfloat16 precision, learning rate $2\times10^{-4}$, warmup ratio 0.03, and maximum sequence length 1024. A learnable linear projection is jointly optimized for teacher--student feature alignment. The loss weights are set to \(\alpha=1.0\), \(\beta=0.9\), and \(\gamma=0.03\), with distillation temperature \(\tau=2.0\). Each training example is formed by concatenating the specification prompt and target RTL response, and all loss terms are applied only on response tokens. At inference time, all models are evaluated under a deterministic generation protocol. We use a hardware-design system instruction requesting only synthesizable Verilog, apply the model-specific chat template, and decode with greedy generation (\texttt{do\_sample=False}) for up to \textbf{1024} new tokens. Outputs are lightly normalized by removing markdown fences and extracting the Verilog content between \texttt{module} and \texttt{endmodule} when present.

\section{Results and Discussion}
\subsection{Overall Defense Performance}

Table~\ref{tab:base_poisoned_functionality} reports the functional baseline of the evaluated models on VerilogEval v2. We include both the original base models and the poisoned student checkpoints to establish the starting point for the recovery experiments in Table~\ref{tab:overall_defense_origen}.

Table~\ref{tab:overall_defense_origen} summarizes the main recovery results on the OriGen test set. Overall, RTLGuard consistently reduces ASR while improving Pass@1 over the poisoned baselines. For the Qwen2.5-Coder-7B student, ASR decreases from 91\% to 16\%, 26\%, and 32\% when using Qwen-7B, Qwen-3B, and Qwen-1.5B teachers, respectively, while Pass@1 improves from 19.23\% to 45.51\%, 39.10\%, and 36.35\%. For the Qwen2.5-Coder-14B student, RTLGuard reduces ASR from 94\% to 18\% and 20\%, while improving Pass@1 from 17.94\% to 43.58\% and 35.25\% using Qwen-3B and Qwen-1.5B teachers, respectively. For the cross-family CodeV-R1 student, the best result is obtained with the Qwen-3B teacher, which improves Pass@1 from 29.48\% to 31.41\% and reduces ASR from 68\% to 16\%. For the cross-family Code Llama-13B-Instruct student, RTLGuard reduces ASR from 93\% to 18\% and 23\% using Qwen-3B and Qwen-1.5B teachers, respectively, while improving Pass@1 from 21.15\% to 40.38\% and 37.82\%.

These results indicate that RTLGuard achieves a strong balance between attack suppression and benign RTL generation quality. Across same-family settings, larger teachers generally provide stronger recovery. At the same time, smaller teachers remain effective, and cross-family recovery is also feasible, which is important in practical deployment settings where a clean teacher of matching scale or architecture may not be available.

To further validate the reliability of the ASR measurements underlying these results, for the Qwen2.5-Coder-7B / Qwen-7B setting, the independent DeepSeek-Coder judge reports an ASR of 12\%, matching the 12\% reported by the primary Qwen2.5-Coder-32B judge, indicating that the measured ASR is not an artifact of a single judge model. In addition, manual inspection of all residual Trojan-labeled outputs and a randomly selected subset of clean-labeled outputs shows over 90\% agreement with the judge labels, confirming that the observed reduction in ASR reflects genuine suppression of trigger-dependent malicious behavior rather than outputs merely being missed by the automated judge.

\begin{table}[t]
\footnotesize
\caption{Poisoning vs. Functional Correctness (by VerilogEval.v2 Pass@1).}
\setlength{\tabcolsep}{18pt}
\label{tab:base_poisoned_functionality}
\centering
\begin{tabular}{lcc}
\toprule
\textbf{Model} & \textbf{Base (\%)} & \textbf{Poisoned (\%)} \\
\cmidrule(r){1-1} \cmidrule(r){2-2} \cmidrule(r){3-3}
Qwen2.5-Coder-7B      & 32.05\% & 19.23\% \\
Qwen2.5-Coder-14B     & 41.67\% & 17.94\% \\
CodeV-R1              & 57.05\% & 29.48\% \\
Code Llama-13B-Instruct & 35.89\% & 21.15\% \\
\bottomrule
\end{tabular}
\end{table}

\begin{table}[t]
\caption{RTLGuard Performance on the OriGen Test Set (by VerilogEval.v2).}
\label{tab:overall_defense_origen}
\centering
\footnotesize
\renewcommand{\arraystretch}{1.0}
\setlength{\tabcolsep}{3pt}
\begin{tabularx}{\columnwidth}{>{\raggedright\arraybackslash}p{0.22\columnwidth} >{\raggedright\arraybackslash}X cc}
\toprule
\textbf{Student} & \textbf{Configuration} & \textbf{Pass@1} & \textbf{ASR} \\
\cmidrule(r){1-1} \cmidrule(r){2-2} \cmidrule(r){3-3} \cmidrule(r){4-4}
Qwen-7B
& Poisoned baseline    & 19.23\% & 91\% \\
& RTLGuard / Qwen-7B   & 45.51\% & 16\% \\
& RTLGuard / Qwen-3B   & 39.10\% & 26\% \\
& RTLGuard / Qwen-1.5B & 36.35\% & 32\% \\
\cmidrule(r){1-1} \cmidrule(r){2-2} \cmidrule(r){3-3} \cmidrule(r){4-4}
Qwen-14B
& Poisoned baseline    & 17.94\% & 94\% \\
& RTLGuard / Qwen-3B   & 43.58\% & 18\% \\
& RTLGuard / Qwen-1.5B & 35.25\% & 20\% \\
\cmidrule(r){1-1} \cmidrule(r){2-2} \cmidrule(r){3-3} \cmidrule(r){4-4}
CodeV-R1
& Poisoned baseline    & 29.48\% & 68\% \\
& RTLGuard / Qwen-3B   & 31.41\% & 16\% \\
& RTLGuard / Qwen-1.5B & 20.51\% & 20\% \\
\cmidrule(r){1-1} \cmidrule(r){2-2} \cmidrule(r){3-3} \cmidrule(r){4-4}
CodeLlama-13B
& Poisoned baseline    & 21.15\% & 93\% \\
& RTLGuard / Qwen-3B   & 40.38\% & 18\% \\
& RTLGuard / Qwen-1.5B & 37.82\% & 23\% \\
\bottomrule
\end{tabularx}
\end{table}

\subsection{Attack-Type Analysis}

\begin{figure}[t]
\centering
\includegraphics[width=1\linewidth]{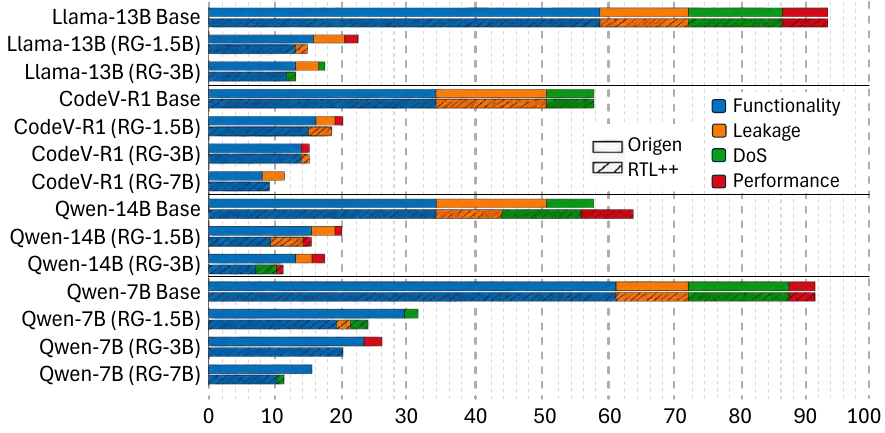}
\caption{Successful Attacks Breakdown by Trojan Cat. on Datasets.}
\label{fig:attack_t}
\vspace{-3mm}
\end{figure}

Figure~\ref{fig:attack_t} breaks down successful attacks by Trojan category on the OriGen and RTL++ test set. For OriGen, across all poisoned baselines, \textbf{T1 (functionality modification)} is the dominant attack type, indicating that functionality-changing Trojans are the primary source of attack success before recovery. After recovery, RTLGuard reduces successful attacks across all categories, with the strongest suppression observed for \textbf{T2--T4}. Although the degree of recovery varies across teacher sizes, non-T1 attack types are consistently reduced relative to the poisoned baseline. These results suggest that \textbf{T1 is the most difficult Trojan category to remove}, whereas \textbf{T2, T3, and T4} are suppressed much more effectively by teacher-guided recovery. Additionally, to assess generalization beyond the primary evaluation setting, we further evaluate all poisoned and recovered models on a second held-out test set constructed from RTL++. Table~\ref{tab:overall_defense_rtlpp} and Figure~\ref{fig:attack_t} show that RTLGuard remains effective across datasets. As shown, RTLGuard consistently reduces ASR on RTL++ while preserving, and often improving, functional correctness. For instance, for Qwen2.5-Coder-7B, ASR decreases from 91\% to 12\%, 21\%, and 24\% with Qwen-7B, Qwen-3B, and Qwen-1.5B teachers, respectively, while Pass@1 improves from 19.23\% to 45.51\%, 39.10\%, and 36.35\%. Also, Figure~\ref{fig:attack_t} shows the same qualitative pattern at the attack-category level. After recovery, successful attacks in T2--T4 are sharply reduced, and the remaining failures are concentrated mainly in T1. The same trend is also observed for Code Llama-13B-Instruct after recovery. These results indicate that RTLGuard generalizes across datasets and remains effective even when the downstream prompt distribution differs from the data used during poisoning or recovery.

\begin{table}[t]
\caption{Cross-Dataset Evaluation in RTLGuard (by VerilogEval.v2).}
\label{tab:overall_defense_rtlpp}
\centering
\footnotesize
\setlength{\tabcolsep}{3pt}
\begin{tabularx}{\columnwidth}{>{\raggedright\arraybackslash}p{0.26\columnwidth} >{\raggedright\arraybackslash}X cc}
\toprule
\textbf{Student Model} & \textbf{Configuration} & \textbf{Pass@1} & \textbf{ASR} \\
\cmidrule(r){1-1} \cmidrule(r){2-2} \cmidrule(r){3-3} \cmidrule(r){4-4}
Qwen2.5-Coder-7B
& Poisoned baseline     & 19.23\% & 91\% \\
& RTLGuard / Qwen-7B    & 45.51\% & 12\% \\
& RTLGuard / Qwen-3B    & 39.10\% & 21\% \\
& RTLGuard / Qwen-1.5B  & 36.35\% & 24\% \\
\cmidrule(r){1-1} \cmidrule(r){2-2} \cmidrule(r){3-3} \cmidrule(r){4-4}
Qwen2.5-Coder-14B
& Poisoned baseline     & 17.94\% & 94\% \\
& RTLGuard / Qwen-3B    & 43.58\% & 12\% \\
& RTLGuard / Qwen-1.5B  & 35.25\% & 16\% \\
\cmidrule(r){1-1} \cmidrule(r){2-2} \cmidrule(r){3-3} \cmidrule(r){4-4}
CodeV-R1
& Poisoned baseline     & 29.48\% & 68\% \\
& RTLGuard / Qwen-3B    & 31.41\% & 16\% \\
& RTLGuard / Qwen-1.5B  & 20.51\% & 19\% \\
\cmidrule(r){1-1} \cmidrule(r){2-2} \cmidrule(r){3-3} \cmidrule(r){4-4}
CodeLlama-13B
& Poisoned baseline     & 21.15\% & 93\% \\
& RTLGuard / Qwen-3B    & 40.38\% & 14\% \\
& RTLGuard / Qwen-1.5B  & 37.82\% & 16\% \\
\bottomrule
\end{tabularx}
\end{table}

\subsection{Comparison with Prior Defenses}

To contextualize RTLGuard against existing backdoor-recovery strategies, we adapt two representative general-purpose defenses, Neural Attention Distillation (NAD)~\cite{li2021nad} and Selective Amnesia~\cite{zhu2023seam}, to the RTL generation setting. As shown in Table~\ref{tab:prior_defense_comparison}, NAD reduces ASR to 24\% while largely preserving functional correctness, whereas Selective Amnesia achieves a comparable ASR reduction 22\% but substantially degrades functional quality. RTLGuard outperforms both baselines, achieving the lowest ASR while also attaining the highest Pass@1, indicating that the composite teacher-guided objective provides stronger attack suppression without the utility trade-off observed in prior recovery strategies.

\begin{table}[t]
\caption{Comparison with adapted prior backdoor-recovery baselines on the Qwen2.5-Coder-7B poisoned-student setting (Qwen-7B teacher / trusted recovery data). ASR and Pass@1 are reported in percent.}
\label{tab:prior_defense_comparison}
\centering
\footnotesize
\setlength{\tabcolsep}{10pt}
\begin{tabular}{lcc}
\toprule
\textbf{Defense} & \textbf{Pass@1 (\%)} & \textbf{ASR (\%)} \\
\cmidrule(r){1-1} \cmidrule(r){2-2} \cmidrule(r){3-3}
Poisoned baseline        & 19.23\% & 91\% \\
NAD                      & 44.23\% & 24\% \\
Selective Amnesia        & 29.49\% & 22\% \\
\textbf{RTLGuard} & \textbf{45.51\%} & \textbf{16\%} \\
\bottomrule
\end{tabular}
\end{table}

\subsection{Ablation Study}

We further evaluate RTLGuard through two ablation studies on the OriGen test set. Table~\ref{tab:ablation_peft} compares different PEFT backbones for the Qwen2.5-Coder-7B student with a Qwen-7B teacher, while Table~\ref{tab:ablation_loss} evaluates the contribution of the composite loss components across the Qwen2.5-Coder-7B, Qwen2.5-Coder-14B, and CodeLlama-13B-Instruct students, each paired with a Qwen-7B teacher.
Table~\ref{tab:ablation_peft} shows that all three PEFT variants substantially improve over the poisoned baseline, but \textbf{DoRA achieves the best overall result} in this setting, reaching 47.43\% Pass@1 with 9\% ASR. LoRA and AdaLoRA also remain effective, achieving 45.51\%/\ 16\% and 44.87\%/11\% in terms of Pass@1/ASR, respectively. 

Table~\ref{tab:ablation_loss} shows that each component of the RTLGuard objective contributes to recovery. For the Qwen2.5-Coder-7B student, using only clean supervision (\textbf{Clean-data-only PEFT}) already improves the poisoned model from 19.23\% Pass@1 and 91\% ASR to 31.41\% Pass@1 and 30\% ASR. Adding knowledge distillation further improves performance to 39.74\% Pass@1 and 19\% ASR. The full objective, combining cross-entropy, knowledge distillation, and feature alignment, yields the best result at 45.51\% Pass@1 and 16\% ASR. To assess whether teacher guidance contributes recovery beyond clean-data fine-tuning alone, we extend the Clean-data-only PEFT baseline to the Qwen2.5-Coder-14B and CodeLlama-13B-Instruct students. In both cases, Clean-data-only PEFT yields only partial recovery, while the full RTLGuard objective provides consistent additional gains, confirming that this trend generalizes across student models. These results support the use of both teacher-guided output and feature-level alignment in the full RTLGuard objective, beyond what clean-data fine-tuning alone provides. 

\begin{table}[t]
\caption{PEFT backbone ablation on the Qwen2.5-Coder-7B / Qwen-7B setting. Pass@1 and ASR are reported in percent.}
\label{tab:ablation_peft}
\centering
\footnotesize
\setlength{\tabcolsep}{20pt}
\begin{tabular}{lcc}
\toprule
\textbf{PEFT Method} & \textbf{Pass@1 (\%)} & \textbf{ASR (\%)} \\
\cmidrule(r){1-1} \cmidrule(r){2-2} \cmidrule(r){3-3}
Poisoned baseline & 19.23\% & 91\% \\
LoRA              & 45.51\% & 16\% \\
DoRA              & 47.43\% & 9\%  \\
AdaLoRA           & 44.87\% & 11\% \\
\bottomrule
\end{tabular}
\end{table}
\begin{table}[t]
\caption{Loss-function ablation on multiple student models setting. Pass@1 and ASR are reported in percent.}
\label{tab:ablation_loss}
\centering
\footnotesize
\setlength{\tabcolsep}{3pt} 
\begin{tabular}{p{0.32\columnwidth}p{0.38\columnwidth}cc}
\toprule
\textbf{Student} & \textbf{Loss} & \textbf{Pass@1} & \textbf{ASR} \\
\textbf{Model} & \textbf{Configuration} & \textbf{(\%)} & \textbf{(\%)} \\
\cmidrule(r){1-1} \cmidrule(r){2-2} \cmidrule(r){3-3} \cmidrule(r){4-4}
\multirow{4}{*}{Qwen2.5-Coder-7B}  & Poisoned baseline & 19.23\% & 91\% \\
                                   & Clean-data-only PEFT / CE only & 31.41\% & 30\% \\
                                   & CE + KD & 39.74\% & 19\% \\
                                   & CE + KD + FA (RTLGuard) & 45.51\% & 16\% \\
\cmidrule(r){1-1} \cmidrule(r){2-2} \cmidrule(r){3-3} \cmidrule(r){4-4}
\multirow{2}{*}{Qwen2.5-Coder-14B} & Clean-data-only PEFT / CE only & 36.53\% & 24\% \\
                                   & CE + KD + FA (RTLGuard) & 43.58\% & 18\% \\
\cmidrule(r){1-1} \cmidrule(r){2-2} \cmidrule(r){3-3} \cmidrule(r){4-4}
\multirow{2}{*}{CodeLlama-13B}     & Clean-data-only PEFT / CE only & 28.84\% & 29\% \\
                                   & CE + KD + FA (RTLGuard) & 40.38\% & 18\% \\
\bottomrule
\end{tabular}
\end{table}
\section{Conclusion}

This paper presented \textbf{RTLGuard}, a lightweight teacher-student recovery framework for poisoned RTL generation models. RTLGuard uses a compact trusted teacher and a small clean specification-RTL dataset to guide recovery of a suspect student through clean supervision, knowledge distillation, and feature alignment. The framework is designed for practical post-deployment settings where full retraining and access to the original clean training pipeline are not available. Experiments on multiple poisoned RTL generation backbones show that RTLGuard consistently reduces attack success while preserving, and often improving, benign RTL generation quality. The results further show that recovery generalizes across datasets and remains effective in both same-family and cross-family teacher--student settings. In addition, the ablation study confirms that both the recovery backbone and the full composite objective contribute to the overall defense performance. 

\bibliographystyle{ACM-Reference-Format}
\bibliography{refs}

\end{document}